\documentclass[british,a4paper,DIV=12,headings=standardclasses]{scrartcl}
\PassOptionsToPackage{natbib=true}{biblatex}
\usepackage[T1]{fontenc}
\usepackage[utf8]{inputenc}
\usepackage{color}
\usepackage{babel}
\usepackage{refstyle}
\usepackage{float}
\usepackage{booktabs}
\usepackage{amssymb}
\usepackage{graphicx}
\usepackage{microtype}
\usepackage[pdfusetitle,
 bookmarks=true,bookmarksnumbered=false,bookmarksopen=false,
 breaklinks=true,pdfborder={0 0 0},pdfborderstyle={},backref=false,colorlinks=true]
 {hyperref}
\hypersetup{
 ocgcolorlinks,linkcolor={dark-red},citecolor={dark-blue},urlcolor={blue}}

\makeatletter

\providecommand{\tabularnewline}{\\}
\newcommand{\lyxdot}{.}

\usepackage{paralist}
\@ifundefined{lettrine}{\usepackage{lettrine}}{}

\usepackage[rm={osf,tab}]{cfr-lm}
  \pdfmapfile{-lm.map}
  \pdfmapfile{-clm.map}
  \pdfmapfile{+mlm-substitute.map} % from mlmodern
  \pdfmapfile{+cmlm-substitute.map}
  \pdfmapline{-lmsy5<lm-mathsy.enc<mlmsy5.pfb}
  \pdfmapline{-lmsy6<lm-mathsy.enc<mlmsy6.pfb}
  \pdfmapline{-lmsy7<lm-mathsy.enc<mlmsy7.pfb}
  \pdfmapline{-lmsy8<lm-mathsy.enc<mlmsy8.pfb}
  \pdfmapline{-lmsy9<lm-mathsy.enc<mlmsy9.pfb}
  \pdfmapline{-lmsy10<lm-mathsy.enc<mlmsy10.pfb}
  \pdfmapline{-lmbsy5<lm-mathsy.enc<mlmbsy5.pfb}
  \pdfmapline{-lmbsy7<lm-mathsy.enc<mlmbsy7.pfb}
  \pdfmapline{-lmbsy10<lm-mathsy.enc<mlmbsy10.pfb}
  \pdfmapfile{+mlmsy.map}
  \SetSymbolFont{symbols}{normal}{OMS}{mlmsy}{m}{n}
  \SetSymbolFont{symbols}{bold}  {OMS}{mlmsy}{b}{n}
\usepackage{MnSymbol}
\newcommand*\mlm@mnscale{1.12527}
\DeclareFontShape{U}{MnSymbolA}{m}{n}{
    <-6>  s*[\mlm@mnscale] MnSymbolA5
   <6-7>  s*[\mlm@mnscale] MnSymbolA6
   <7-8>  s*[\mlm@mnscale] MnSymbolA7
   <8-9>  s*[\mlm@mnscale] MnSymbolA8
   <9-10> s*[\mlm@mnscale] MnSymbolA9
  <10-12> s*[\mlm@mnscale] MnSymbolA10
  <12->   s*[\mlm@mnscale] MnSymbolA12}{}
\DeclareFontShape{U}{MnSymbolA}{b}{n}{
    <-6>  s*[\mlm@mnscale] MnSymbolA-Bold5
   <6-7>  s*[\mlm@mnscale] MnSymbolA-Bold6
   <7-8>  s*[\mlm@mnscale] MnSymbolA-Bold7
   <8-9>  s*[\mlm@mnscale] MnSymbolA-Bold8
   <9-10> s*[\mlm@mnscale] MnSymbolA-Bold9
  <10-12> s*[\mlm@mnscale] MnSymbolA-Bold10
  <12->   s*[\mlm@mnscale] MnSymbolA-Bold12}{}
\DeclareFontShape{U}{MnSymbolB}{m}{n}{
    <-6>  s*[\mlm@mnscale] MnSymbolB5
   <6-7>  s*[\mlm@mnscale] MnSymbolB6
   <7-8>  s*[\mlm@mnscale] MnSymbolB7
   <8-9>  s*[\mlm@mnscale] MnSymbolB8
   <9-10> s*[\mlm@mnscale] MnSymbolB9
  <10-12> s*[\mlm@mnscale] MnSymbolB10
  <12->   s*[\mlm@mnscale] MnSymbolB12}{}
\DeclareFontShape{U}{MnSymbolB}{b}{n}{
    <-6>  s*[\mlm@mnscale] MnSymbolB-Bold5
   <6-7>  s*[\mlm@mnscale] MnSymbolB-Bold6
   <7-8>  s*[\mlm@mnscale] MnSymbolB-Bold7
   <8-9>  s*[\mlm@mnscale] MnSymbolB-Bold8
   <9-10> s*[\mlm@mnscale] MnSymbolB-Bold9
  <10-12> s*[\mlm@mnscale] MnSymbolB-Bold10
  <12->   s*[\mlm@mnscale] MnSymbolB-Bold12}{}
\DeclareFontShape{U}{MnSymbolC}{m}{n}{
    <-6>  s*[\mlm@mnscale] MnSymbolC5
   <6-7>  s*[\mlm@mnscale] MnSymbolC6
   <7-8>  s*[\mlm@mnscale] MnSymbolC7
   <8-9>  s*[\mlm@mnscale] MnSymbolC8
   <9-10> s*[\mlm@mnscale] MnSymbolC9
  <10-12> s*[\mlm@mnscale] MnSymbolC10
  <12->   s*[\mlm@mnscale] MnSymbolC12}{}
\DeclareFontShape{U}{MnSymbolC}{b}{n}{
    <-6>  s*[\mlm@mnscale] MnSymbolC-Bold5
   <6-7>  s*[\mlm@mnscale] MnSymbolC-Bold6
   <7-8>  s*[\mlm@mnscale] MnSymbolC-Bold7
   <8-9>  s*[\mlm@mnscale] MnSymbolC-Bold8
   <9-10> s*[\mlm@mnscale] MnSymbolC-Bold9
  <10-12> s*[\mlm@mnscale] MnSymbolC-Bold10
  <12->   s*[\mlm@mnscale] MnSymbolC-Bold12}{}
\DeclareFontShape{U}{MnSymbolD}{m}{n}{
    <-6>  s*[\mlm@mnscale] MnSymbolD5
   <6-7>  s*[\mlm@mnscale] MnSymbolD6
   <7-8>  s*[\mlm@mnscale] MnSymbolD7
   <8-9>  s*[\mlm@mnscale] MnSymbolD8
   <9-10> s*[\mlm@mnscale] MnSymbolD9
  <10-12> s*[\mlm@mnscale] MnSymbolD10
  <12->   s*[\mlm@mnscale] MnSymbolD12}{}
\DeclareFontShape{U}{MnSymbolD}{b}{n}{
    <-6>  s*[\mlm@mnscale] MnSymbolD-Bold5
   <6-7>  s*[\mlm@mnscale] MnSymbolD-Bold6
   <7-8>  s*[\mlm@mnscale] MnSymbolD-Bold7
   <8-9>  s*[\mlm@mnscale] MnSymbolD-Bold8
   <9-10> s*[\mlm@mnscale] MnSymbolD-Bold9
  <10-12> s*[\mlm@mnscale] MnSymbolD-Bold10
  <12->   s*[\mlm@mnscale] MnSymbolD-Bold12}{}

\g@addto@macro\bfseries{\boldmath}
\usepackage{xcolor}
  \definecolor{dark-red}{rgb}{0.6,0,0}
  \definecolor{dark-blue}{rgb}{0,0.0,0.6}
  \definecolor{blue}{rgb}{0,0,0.8}

\usepackage{caption}
\usepackage[font=small,labelfont=bf]{caption}

\DeclareTextFontCommand{\texttt}{\ttfamily\fontseries{l}\selectfont}

\makeatother

\usepackage[maxnames=99]{biblatex}
\begin{document}
\title{\fontseries{n}\selectfont\mathversion{normal} Who decides the norms
in TCS? Analysing the composition of the STOC 2027 committees and
of the Simons Institute “AI$+$TCS” working group}
\author{Guillaume Munch-Maccagnoni\thanks{INRIA. Email: \protect\href{mailto:Guillaume.Munch-Maccagnoni@inria.fr}{Guillaume.Munch-Maccagnoni@inria.fr}.
An anonymous team of volunteer Ph.D. students has contributed to the
data collection, as described on page~\pageref{par:Methodology-and-LLM}.}}
\date{23rd November 2026}

\maketitle
\noindent\hspace*{\fill}\rule[-1ex]{0.88\textwidth}{0.3pt}\hspace*{\fill}
\begin{abstract}
\noindent\textbf{Summary}\enspace{}A working group organised by
the Simons Institute released a report last September containing recommendations
aiming to increase the usage of LLMs in theoretical computer science
(TCS), including introducing them for communication and evaluation.
Most of these recommendations were enacted by the STOC\footnote{The Symposium on Theory of Computing of the ACM (Association for Computing
Machinery).} 2027 programme committee and the ACM SIGACT\footnote{The Special Interest Group on Algorithms and Computation Theory of
the ACM.} executive committee (as the steering committee of STOC). The compositions
of the Simons Institute working group and of the STOC programme and
steering committees are analysed as follows.~
\begin{inparaitem}
\item The composition of the Simons Institute working group is analysed
along two axes: country of affiliation, and ties with industry (i.e.
affiliated with or funded by). The results (on page~\pageref{fig:simons2})
show an overwhelming representation of researchers from US institutions
and private companies, and a large over-representation of researchers
with ties to AI/Big Tech companies (in the present or the recent past).~%{\small\par}
\item The composition of the STOC 2027 and 2026 programme committees is
analysed in terms of country of affiliation. The results (on page~\pageref{fig:stoc2027})
show a large over-representation in 2027 of researchers from US institutions/companies,
and large under-representation of researchers from EU and Asian institutions,
as compared to 2026.~%{\small\par}
\item The composition of the ACM SIGACT extended executive committee is
analysed along the same two axes as the Simons Institute working group:
country and industrial ties. The results (on page~\pageref{fig:sigact2})
show the same overwhelming representation of researchers from US institutions/companies,
and a large representation of researchers with ties to AI/Big Tech
companies.%{\small\par}
\end{inparaitem}
\end{abstract}
\noindent\hfill{}\rule[2ex]{0.88\textwidth}{0.3pt}\hfill{}

\lettrine{T}{he ACM conference STOC 2027}{,\footnote{\href{https://acm-stoc.org/stoc2027/}{https://acm-stoc.org/stoc2027/}
(All linked web pages have been accessed between Nov. 17-20th 2026.)} one of the top publishing venues in theoretical computer science,
has announced in September 2026 that it would allow LLM use in its
peer-review process, among other moves that came as a surprise to
many colleagues in the broader TCS community. The submission process
for STOC also required that authors post simultaneously their submission
on arXiv, to counteract a standard objection against uploads to LLM
services, namely the confidentiality of submissions.\footnote{This was later amended to requiring consent for the submitted papers
being made public on the STOC webpage, together with a simple recommendation
to place on a preprint repository. The submissions are indeed presently
available at \href{https://sigact-stoc.github.io/stoc2027/}{https://sigact-stoc.github.io/stoc2027/}.} Double-blind reviewing was abandoned. These policies were presented
as \emph{“experimental”}---a word used in this case for changes affecting
everyone which may still need adjustments in the future (but unlikely
to return to the status quo ante), as opposed to controlled changes
based on clear hypotheses and success or failure criteria stated in
advance and involving a degree of choice and consent.}

This move was organised in coordination with a working group called
\emph{“AI and TCS: the Next Six Months”}\footnote{\href{https://simons.berkeley.edu/news-publications-videos/ai-tcs-working-group}{https://simons.berkeley.edu/news-publications-videos/ai-tcs-working-group}}\textsuperscript{,}\footnote{\href{https://simons.berkeley.edu/AITCSreport}{https://simons.berkeley.edu/AITCSreport}
(document with creation date 30/09/2026)} organised by the Simons Institute, a research institute in theoretical
computer science located in Berkeley, California. This working group
has released shortly after the STOC announcement a report that reads
as a playbook to introduce LLM-based technology and services in all
aspects of theoretical computer science---including communication
and evaluation, e.g. writing papers and reviews. As the report notes,
\emph{“the STOC 2027 PC has decided to adopt almost all of our guidelines”}
before it was even made public. It contains such recommendations for
conferences as, indeed, requiring arXiv postings with an explicit
goal of permitting LLM usage by reviewers. Yet the report is devoid
of any reference to a scientific literature supporting the idea of
using LLMs in peer review. It also contains other recommendations
with no scientific basis in the capabilities of LLMs, such as \emph{“Tracing
the Lineage of AI-Generated Ideas”}, or that are empty, such as requiring
an attestation from \emph{“the authors certifying that they understand
and have verified the work they are submitting”}.

The report calls for a \emph{“coordinated”} action to enact these
new norms in conferences, with a conclusion given the shape of a political
manifesto:
\begin{quote}
\emph{``}{[}...{]} \emph{So what now? The power comes from what happens
next, and that will involve individuals, departments, and institutions
considering these recommendations}~{[}...{]}.

\emph{We stress that this is a moment when }every person in our community
has a role to play.\emph{}~{[}...{]}\emph{ We tried our best to
articulate the path forward on implementation and did so with the
firm belief in the capabilities of our community to collectively solve
the challenges that arise. That belief extends to you --- a fellow
community member, with likely the same concerns and questions that
led us to initiate this work.}~{[}...{]}\emph{''}
\end{quote}
The report draws in particular from the vocabulary of crisis, claiming
that these coordinated actions are necessary due to \emph{“rapid”}
and \emph{“dramatic”} advances in \emph{“AI capabilities”}, to \emph{“provide
clarity in a chaotic time”}. As the report further explains, \emph{``the
actions recommended in this report are framed to be executed as soon
as possible, ideally in the next six months”}.

Such a hurry is employed to justify a skewed composition of the 28-member
group. As the report admits, \emph{“the short timeline and in-person
constraint naturally biased the workshop towards participants who
could come to Berkeley on short notice”}. The Simons Institute is
funded by both public and private sources, including many AI/Big Tech
companies.\footnote{\emph{``We collaborate with industry through our Institutional Partner
program, which currently includes these }partners\emph{: ADIA Lab,
Apple, Bridgewater Associates, Google, Jane Street, and Oski Technology/Nvidia,
and these }sponsors\emph{: Algorand, AWS, NTT Research, and Stellar
Foundation. Former Industry partners and sponsors include Calibra/Facebook,
IBM, IVADO, JP Morgan, Microsoft Research, PayPal, ProtocolAi, RelationalAi,
Ripple, Roc360, Qualcomm, Swiss Re, and VMware.'' }(\href{https://simons.berkeley.edu/support/funders}{https://simons.berkeley.edu/support/funders})} By AI company we mean a private company operating significantly in
AI, and by Big Tech we mean the largest technology corporations, which
are also largely invested in AI.

\paragraph*{Composition of the AI$+$TCS working group by country of affiliation.}

\begin{figure}
\begin{centering}
\includegraphics[scale=0.7]{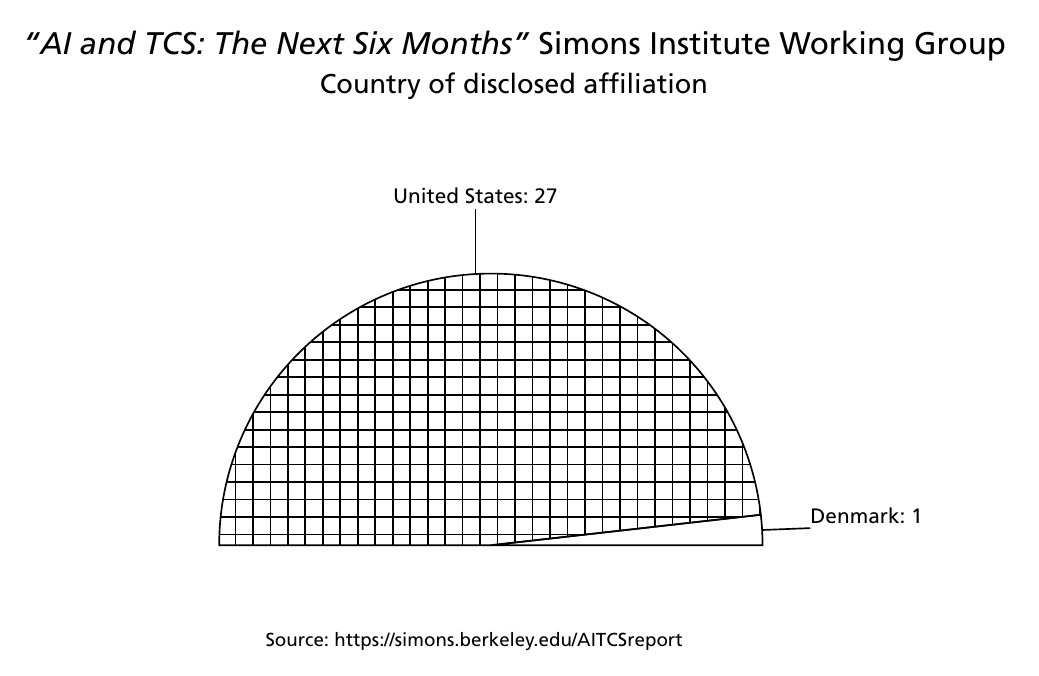}
\par\end{centering}
\caption{\emph{Alt text:} Pie chart entitled “AI and TCS: The Next Six Months”
Simons Institute Working Group --- Country of disclosed affiliation.
Data: United States (27), Denmark (1). Source: \protect\href{https://simons.berkeley.edu/AITCSreport}{https://simons.berkeley.edu/AITCSreport}.}\label{fig:simons2}

\bigskip{}

\begin{centering}
\includegraphics[scale=0.7]{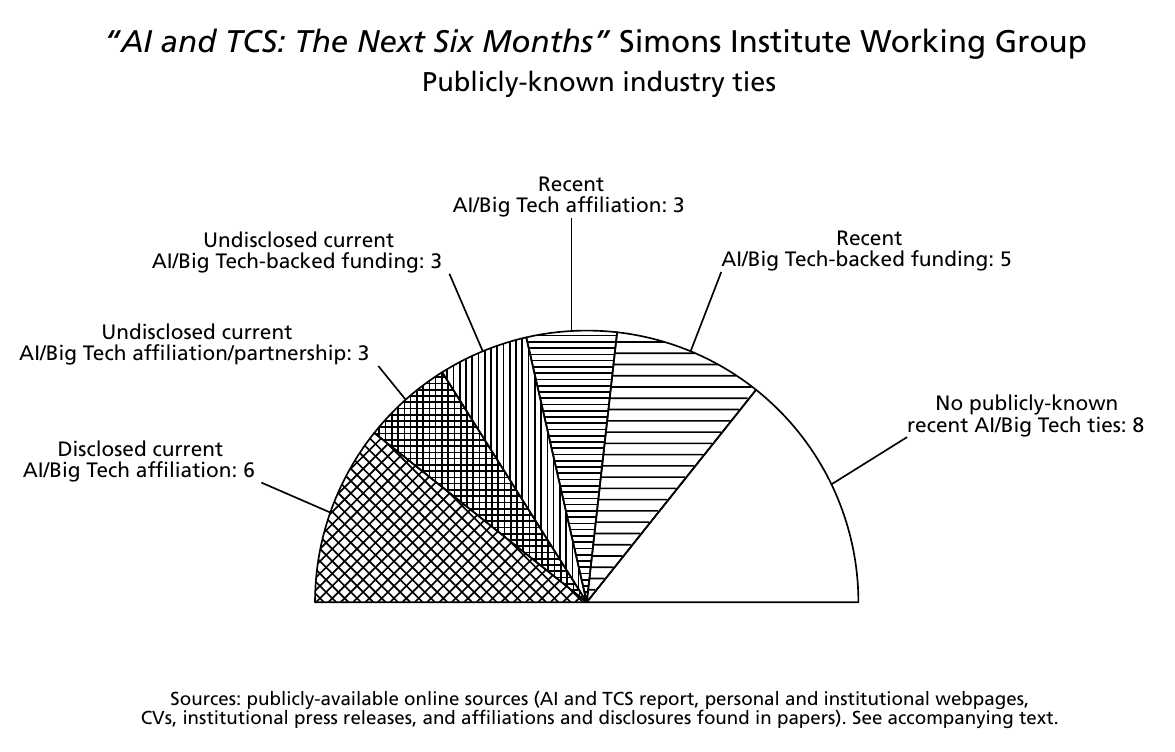}
\par\end{centering}
\caption{\emph{Alt text:} Pie chart entitled “AI and TCS: The Next Six Months”
Simons Institute Working Group --- Publicly-known industry ties.
Data: Disclosed current AI/Big Tech affiliation (6), Undisclosed current
AI/Big Tech affiliation/partnership (3), Undisclosed current AI/Big
Tech-backed funding (3), Recent AI/Big Tech affiliation (3), Recent
AI/Big Tech-backed funding (5), No publicly-known recent AI/Big Tech
ties (8). Source: publicly-available online sources (AI$+$TCS report,
personal and institutional webpages, CVs, institutional press releases,
and affiliations and disclosures found in papers). See accompanying
text.}\label{fig:simons}
\end{figure}

Since the report claims to derive its legitimacy from a \emph{“broad
consensus from the group”}, and that \emph{“the organizers did their
best to gather a diverse sample of the TCS community”}, it is interesting
to study its composition. The chart in \figref{simons2} shows the
proportion of US vs non-US participants in terms of disclosed affiliation,
also available as a table in the appendix. The data shows an overwhelming
representation of researchers from US institutions/companies. The
working group includes 7 women (25.0\%).

\paragraph*{Composition of the STOC 2027 and 2026 programme committees by country
of affiliation and gender.}

\begin{figure}
\begin{centering}
\includegraphics[scale=0.7]{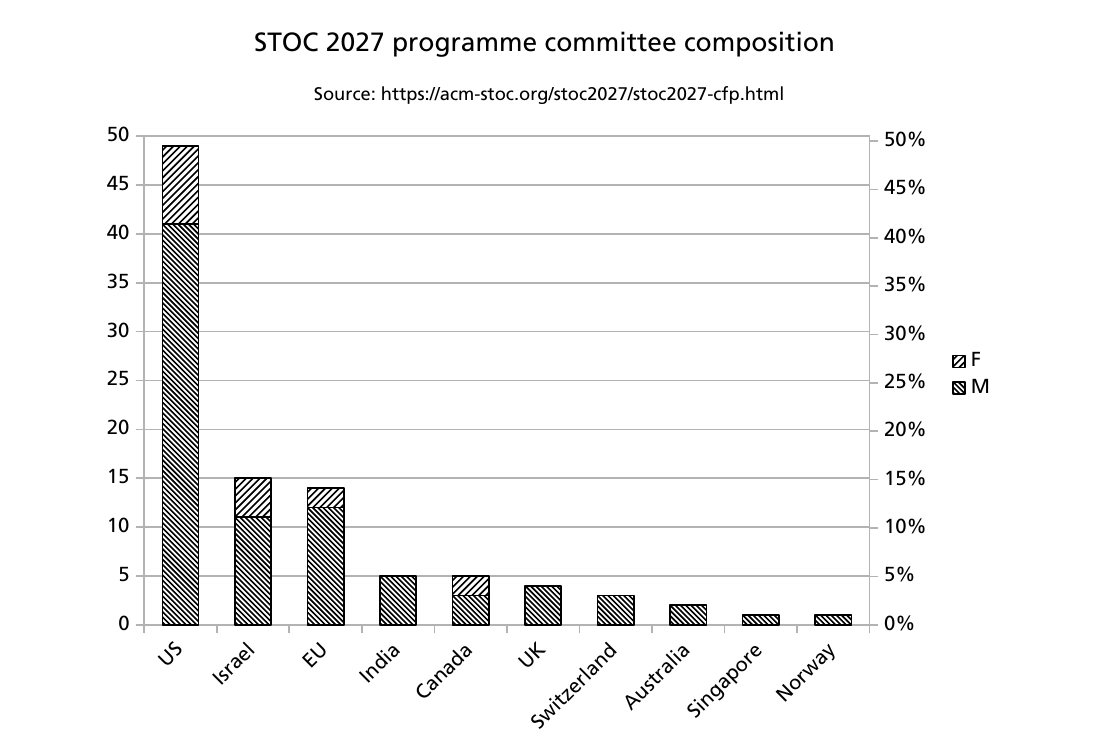}
\par\end{centering}
\begin{centering}
\caption{\emph{Alt text:} Bar chart entitled STOC 2027 programme committee
composition. Data (abridged): US (49.5\%), Israel (15.2\%), EU (14.1\%),
India (5.1\%), Canada (5.1\%), UK (4.0\%), Switzerland (3.0\%), Australia
(2.0\%), Singapore (1.0\%), Norway (1.0\%). Source: \protect\href{https://acm-stoc.org/stoc2027/stoc2027-cfp.html}{https://acm-stoc.org/stoc2027/stoc2027-cfp.html}.}\label{fig:stoc2027}
\bigskip{}
\par\end{centering}
\centering{}\includegraphics[scale=0.7]{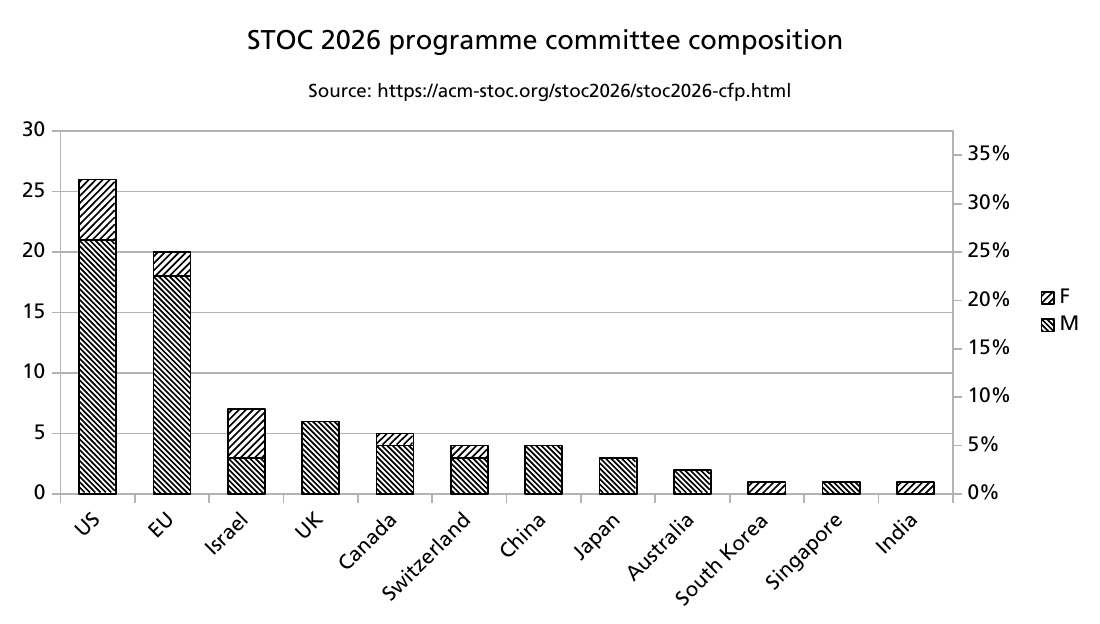}\caption{\emph{Alt text:} Bar chart entitled STOC 2026 programme committee
composition. Data (abridged): US (32.5\%), EU (25.0\%), Israel (8.8\%),
UK (7.5\%), Canada (6.3\%), Switzerland (5.0\%), China (5.0\%), Japan
(3.8\%), Australia (2.5\%), South Korea (1.3\%), Singapore (1.3\%),
India (1.3\%). Source: \protect\href{https://acm-stoc.org/stoc2026/stoc2026-cfp.html}{https://acm-stoc.org/stoc2026/stoc2026-cfp.html}.}\label{fig:stoc2026}
\end{figure}

\noindent Next we look similarly at the composition of the STOC 2027
programme committee (99 members), as compared to the STOC 2026 programme
committee (80 members). This is relevant and interesting because the
policy changes have been decided by votes of the programme committee,\footnote{According to a private communication from a member of the STOC 2027
PC.} before final decision by the ACM SIGACT executive committee, which
acts as the steering committee of STOC.\footnote{\href{https://acm-stoc.org/stoc2027/stoc2027-policy-explanations.pdf}{https://acm-stoc.org/stoc2027/stoc2027-policy-explanations.pdf}}
The compositions in terms of country of disclosed affiliation and
gender\footnote{Using declared gender or public pronouns (when available) or inferring
from first names. As no one was found to explicitly identify other
than with ``he'' or ``she'', the charts only show two genders.} are shown in figures~\ref{fig:stoc2027} and \ref{fig:stoc2026}.
The vertical scales are chosen so that they are comparable for the
percentages (i.e. the total sums of bar heights are the same). The
data is also reproduced in tables in the appendix. The data shows
a large over-representation in 2027 of researchers from US institutions/companies
as compared to 2026 (49.5\% vs. 32.5\%), and large under-representation
of researchers from institutions in EU (14.3\% vs. 25\%) and Asia\footnote{South and East Asia. In this instance: China, India, Japan, Singapore,
and South Korea.} (6.1\% vs. 12.5\%) as compared to 2026. The proportion of women is
low and decreased slightly (16.2\% vs. 18.8\%).

\paragraph{Composition of the STOC 2027 steering committee by country of affiliation.}

\begin{figure}
\begin{centering}
\includegraphics[scale=0.7]{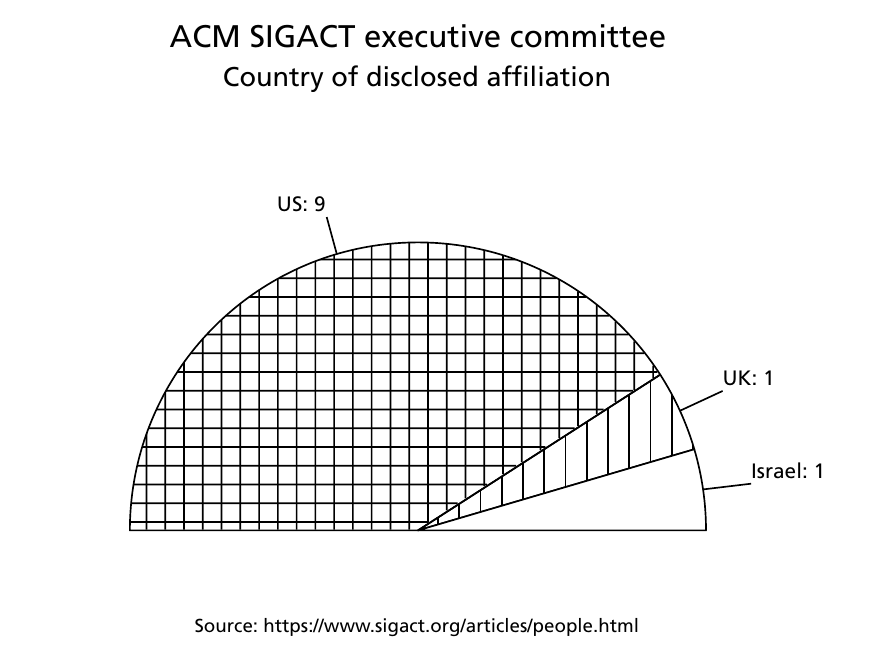}
\par\end{centering}
\caption{\emph{Alt text:} Pie chart entitled ACM SIGACT executive committee
--- Country of disclosed affiliation. Data: United States (9), United
Kingdom (1), Israel (1). Source: \protect\href{https://www.sigact.org/articles/people.html}{https://www.sigact.org/articles/people.html}.}\label{fig:sigact2}

\bigskip{}

\begin{centering}
\includegraphics[scale=0.7]{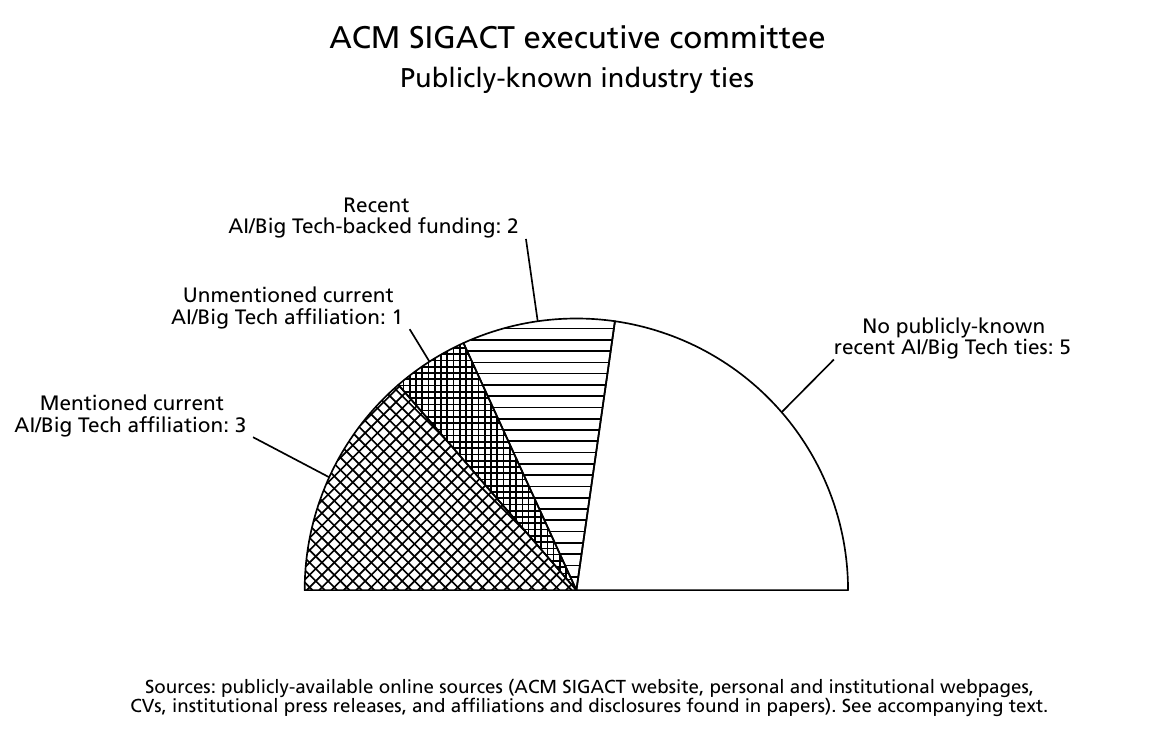}
\par\end{centering}
\caption{\emph{Alt text:} Pie chart entitled ACM SIGACT executive committee
--- Publicly-known industry ties. Data: Mentioned current AI/Big
Tech affiliation (3), Unmentioned current AI/Big Tech affiliation
(1), Recent AI/Big Tech-backed funding (2), No publicly-known recent
AI/Big Tech ties (5). Source: publicly-available online sources (ACM
SIGACT website, personal and institutional webpages, CVs, institutional
press releases, and affiliations and disclosures found in papers).
See accompanying text.}\label{fig:sigact}
\end{figure}

\noindent Next we look similarly at the composition of the ACM SIGACT
extended executive committee (11 members). This committee validated
some of the decisions of the STOC 2027 programme committee, whilst
rendering its own decisions for some of the policies.\footnote{According to a private communication from a member of the STOC 2027
PC.} The chart in \figref{sigact2} shows the number and proportion of
members by country of disclosed affiliation, also available as a table
in the appendix. The data shows an overwhelming representation of
researchers from US institutions/companies. The executive committee
includes 4 women (36.4\%).

\paragraph*{Ties with industry within the AI$+$TCS working group.}

\noindent The AI$+$TCS report describes the process of forming the
working group as follows: \emph{“The invitations were sent iteratively,
attempting to balance across seniority, research area, institution,
and gender, and to include recent PC/steering committee members as
well as people active in the public AI/Math/TCS discourse.”} The report
summarises the composition of the group as \emph{``5 graduate students,
3 postdocs, 16 faculty, and 4 people from industry''}. It discloses
6 members affiliated with an AI/Big Tech company.\footnote{This includes one member with a disclosed affiliation to “a16z Crypto”,
which indeed pivoted to AI as it announced its rebranding into “a16z
Frontier” in October 2026. Its parent firm, Andreessen Horowitz, is
also listed by the \href{https://authoritarian-stack.info/}{Authoritarian Stack project}
as a main actor of the US authoritarian tech network.} Neither the potential ties to industry nor the personal and professional
usage of LLM-based technologies and services are listed as criteria
either way. It would be interesting and relevant to look at these
aspects of the working group composition, because the \emph{``broad
consensus''} reported could be affected by phenomena ranging from
selection biases to dependency and vested interests. Also, no process
for screening for potential conflicts of interest is described in
the report.

These questions are, by their very nature, best addressed at the initiative
of the working group organisers. We may nevertheless look specifically
at publicly-known ties to industry of working group members, by which
we mean both industrial affiliations and industrial funding sources,
both current and recent, by using public sources such as personal
and institutional web pages, CVs, affiliations and disclosures found
in papers, etc. (See ``Methodology and LLM disclosure'' further
below.) The results are shown in the chart in \figref{simons}, also
available as a table in the appendix. In addition to 6 members whose
industrial affiliation is disclosed, 3 members have undisclosed strong
ties to AI/Big Tech (two direct affiliations, and an industrial partnership
as co-director of a research institute). Other ties include current
and recent AI/Big Tech-backed grants and awards, and recent AI/Big
Tech affiliations. (When several ties exist, only the most important
one is counted.) All in all, only 8 members had no publicly-known
ties to AI/Big Tech that could be identified. The dataset is not made
public given its nominative nature, and as we consider that it falls
to the organisers of the working group to organise such disclosures.

There is no claim that this data is complete, in fact we have no reason
to believe that it is. Indeed, if the information was often directly
disclosed on personal web pages or CVs, at times the sources were
more indirect. In one instance, some industrial funding could only
be sourced from an institutional press release of a different institution
(that of the co-PI). In another instance, the information for a current
Big Tech affiliation, undisclosed in the report, was also not present
in the CV nor on personal nor institutional web pages. These unknowns
on the exhaustiveness of publicly-available information provide a
second reason to hope that the working group organisers will organise
a disclosure.

\paragraph*{Ties with industry within the STOC 2027 steering committee.}

Similarly, the publicly-known ties to industry of members of the ACM
SIGACT extended executive committee are analysed. The results are
shown in the chart in \figref{sigact}, also available as a table
in the appendix. In addition to 3 members whose affiliation to Big
Tech companies is disclosed on the website of SIGACT\footnote{\href{https://www.sigact.org/articles/people.html}{https://www.sigact.org/articles/people.html}},
there is an additional member with a current Big Tech affiliation\footnote{While unmentioned on the website, this affiliation was disclosed on
candidacy documents for the position.}, and two members who received grants or awards from the industry
in the recent years. (As before, when several ties exist only the
most important one is counted.) One member has strong current and
recent industrial ties that could not be attached to our description
of AI/Big Tech, and only 4 members are without publicly-known ties
to industry (that could be identified).

\paragraph{Concluding remarks.}

The recommendations of the “AI$+$TCS” working group and the actions
of the STOC programme and steering committees can be analysed under
two angles, both in the large representation of US institutions and
Big Tech in norm-setting, and as actions increasing in turn the dependency
on Big Tech. We have previously been warned about the increased dominance
of Big Tech in computer science domains such as natural language processing\footfullcite{Abdalla2023}\edef\abdalla{\number\value{footnote}}
and tech ethics\footfullcite{Abdalla2021}\edef\aabdalla{\number\value{footnote}}.
As techniques and technology requiring large investments become more
relevant to theoretical computer science as instruments, not just
as objects of study, there are indeed risks of an increased dependency
on Big Tech companies.

As Abdalla et al.\footnotemark[\abdalla] recall, such a growing industrial
presence can be beneficial, resulting in greater investment and results.
They also recall the risks, such as centralization of resources, lack
of impartiality, and undue influence, in short the capture of the
field. In the domain of ethics of AI, Abdalla and Abdalla\footnotemark[\aabdalla]
compare the actions of Big Tech companies to the Big Tobacco playbook,
the well-studied actions of the largest tobacco companies that distorted
the public discourse around public health starting from the 1950's
onwards. As computer scientists have been reflecting more and more
about the impact of their field on society, the value of a scientific
discourse free from industrial influence has become increasingly evident. 

In theoretical computer science and mathematics, the news that famous
conjectures have been solved with the use of LLM-based platforms has
certainly created a shock---which indeed the “AI$+$TCS” working
group and the STOC programme and steering committees aimed to (but
also took the opportunity to) react to. This also marks the evolution
of theoretical computer science into a domain in which the same questions
as in the rest of the field ought to be asked: \emph{``who has the
power, what their interests and values might be, and how they are
influencing research.''}\footnotemark[\abdalla]\vspace*{5ex}

\begin{center}
$\divideontimes$\\
$\divideontimes\hspace*{0.8em}\divideontimes$\vspace*{6ex}
\par\end{center}

\paragraph{Methodology and LLM disclosure.}\label{par:Methodology-and-LLM}

LLM-based online chatbots were used to locate and classify relevant
data online, data which was then verified exhaustively by hand. Most
importantly, such an LLM-based service was used to create a first
set of affiliation and funding information pertaining to the Simons
Institute working group. The result was poor as it stood, but it provided
the initial impetus to establish a reliable dataset (the chatbot did
output relevant information that was not disclosed in the report,
and information which would have been missed using a classic search
engine). After a first round of cleanup and verification by hand,
the dataset was formatted suitably for further verification, and dispatched
to a team of volunteer Ph.D. students informed in advance of the origin
of the data, who wished to contribute anonymously. They confirmed
most of the data with the help of online sources, and separately found
relevant data not found with the chatbot. Among the data which they
flagged as unverified, some was eventually validated, some was not
(and thus was removed from the dataset). In this way, each entry was
verified by at least two people.  Due to the amount of work which
the same task would require for the STOC 2026 and 2027 programme committees
(169 individuals), and since we have no a priori reason to believe
that the result would be instructive, the ties to the industry in
the programme committees were not investigated. The present article
is entirely human-thought-out and written.\looseness=-1

\paragraph{Acknowledgements.}

Many thanks to the team of volunteer Ph.D. students, and to other
colleagues for their encouragement and feedback.

\paragraph{License.}

The text as well as the charts can be reused under the conditions
of the license \textsc{\href{https://creativecommons.org/licenses/by-nc-sa/4.0/}{cc by-nc-sa 4.0}}.

%dont show references\printbibliography

\clearpage{}


\clearpage{}

\appendix

\section*{Appendix: Tables}

\begin{table}[H]
\begin{centering}
\begin{tabular}{lr}
\toprule 
\multicolumn{2}{c}{\textbf{AI$+$TCS Country of affiliation}}\tabularnewline
\midrule 
US & 27\tabularnewline
Denmark & 1\tabularnewline
\midrule
\textbf{Total} & \textbf{28}\tabularnewline
\bottomrule
\end{tabular}
\par\end{centering}
\caption{Composition of the AI$+$TCS working group by country of affiliation
(\figref{simons2})}
\end{table}

\begin{table}[H]
\begin{centering}
\begin{tabular}{lr}
\toprule 
\multicolumn{2}{c}{\textbf{AI$+$TCS Ties to industry}}\tabularnewline
\midrule
No publicly-known recent AI/Big Tech ties: & 8\tabularnewline
Recent AI/Big Tech-backed funding: & 5\tabularnewline
Recent AI/Big Tech affiliation: & 3\tabularnewline
Undisclosed current AI/Big Tech-backed funding: & 3\tabularnewline
Undisclosed current AI/Big Tech affiliation/partnership: & 3\tabularnewline
Disclosed current AI/Big Tech affiliation: & 6\tabularnewline
\midrule
\textbf{Total} & \textbf{28}\tabularnewline
\bottomrule
\end{tabular}
\par\end{centering}
\caption{Ties with industry within the AI$+$TCS working group (\figref{simons})}
\end{table}

\begin{table}[H]
\begin{centering}
\begin{tabular}[t]{lrrrrr}
\toprule 
\multicolumn{6}{c}{\textbf{STOC 2027 PC by country of affiliation}}\tabularnewline
\addlinespace[0.5ex]
\textbf{Country} & \textbf{M} & \textbf{F} & \textbf{Tot.} & \textbf{F (\%)} & \textbf{Tot. (\%)}\tabularnewline
\midrule
\addlinespace[0.5ex]
US & 41 & 8 & 49 & 16.3\% & 49.5\%\tabularnewline
\addlinespace[0.5ex]
Israel & 11 & 4 & 15 & 26.7\% & 15.2\%\tabularnewline
\addlinespace[0.5ex]
EU & 12 & 2 & 14 & 14.3\% & 14.1\%\tabularnewline
\addlinespace[0.5ex]
India{*} & 5 & 0 & 5 & 0.0\% & 5.1\%\tabularnewline
\addlinespace[0.5ex]
Canada & 3 & 2 & 5 & 40.0\% & 5.1\%\tabularnewline
\addlinespace[0.5ex]
UK & 4 & 0 & 4 & 0.0\% & 4.0\%\tabularnewline
\addlinespace[0.5ex]
Switzerland & 3 & 0 & 3 & 0.0\% & 3.0\%\tabularnewline
\addlinespace[0.5ex]
Australia & 2 & 0 & 2 & 0.0\% & 2.0\%\tabularnewline
\addlinespace[0.5ex]
Singapore{*} & 1 & 0 & 1 & 0.0\% & 1.0\%\tabularnewline
\addlinespace[0.5ex]
Norway & 1 & 0 & 1 & 0.0\% & 1.0\%\tabularnewline
\midrule
\addlinespace[0.5ex]
{*}Asia (2) & 6 & 0 & 6 & 0.00\% & 6.1\%\tabularnewline
\addlinespace[0.5ex]
\textbf{Total} & \textbf{83} & \textbf{16} & \textbf{99} & \textbf{16.2\%} & \textbf{100.0\%}\tabularnewline
\bottomrule
\addlinespace[0.5ex]
\end{tabular}
\par\end{centering}
\caption{Composition of the STOC 2027 programme committee by country of affiliation
and gender (figure~\ref{fig:stoc2027})}
\end{table}

\begin{table}[H]
\begin{centering}
\begin{tabular}[t]{lrrrrr}
\toprule 
\multicolumn{6}{c}{\textbf{STOC 2026 PC by country of affiliation}}\tabularnewline
\addlinespace[0.5ex]
\textbf{Country} & \textbf{M} & \textbf{F} & \textbf{Tot.} & \textbf{F (\%)} & \textbf{Tot. (\%)}\tabularnewline
\midrule
\addlinespace[0.5ex]
US & 21 & 5 & 26 & 19.2\% & 32.5\%\tabularnewline
\addlinespace[0.5ex]
EU & 18 & 2 & 20 & 10.0\% & 25.0\%\tabularnewline
\addlinespace[0.5ex]
Israel & 3 & 4 & 7 & 57.1\% & 8.8\%\tabularnewline
\addlinespace[0.5ex]
UK & 6 & 0 & 6 & 0.0\% & 7.5\%\tabularnewline
\addlinespace[0.5ex]
Canada & 4 & 1 & 5 & 20.0\% & 6.3\%\tabularnewline
\addlinespace[0.5ex]
Switzerland & 3 & 1 & 4 & 25.0\% & 5.0\%\tabularnewline
\addlinespace[0.5ex]
China{*} & 4 & 0 & 4 & 0.0\% & 5.0\%\tabularnewline
\addlinespace[0.5ex]
Japan{*} & 3 & 0 & 3 & 0.0\% & 3.8\%\tabularnewline
\addlinespace[0.5ex]
Australia & 2 & 0 & 2 & 0.0\% & 2.5\%\tabularnewline
\addlinespace[0.5ex]
South Korea{*} & 0 & 1 & 1 & 100.0\% & 1.3\%\tabularnewline
\addlinespace[0.5ex]
Singapore{*} & 1 & 0 & 1 & 0.0\% & 1.3\%\tabularnewline
\addlinespace[0.5ex]
India{*} & 0 & 1 & 1 & 100.0\% & 1.3\%\tabularnewline
\midrule
\addlinespace[0.5ex]
{*}Asia (5) & 8 & 2 & 10 & 20.00\% & 12.5\%\tabularnewline
\addlinespace[0.5ex]
\textbf{Total} & \textbf{65} & \textbf{15} & \textbf{80} & \textbf{18.8\%} & \textbf{100.0\%}\tabularnewline
\bottomrule
\addlinespace[0.5ex]
\end{tabular}
\par\end{centering}
\caption{Composition of the STOC 2026 programme committee by country of affiliation
and gender (figure~\ref{fig:stoc2026})}
\end{table}

\begin{table}[H]
\begin{centering}
\begin{tabular}{lr}
\toprule 
\multicolumn{2}{c}{\textbf{SIGACT EC Country of affiliation}}\tabularnewline
\midrule 
US & 9\tabularnewline
UK & 1\tabularnewline
Israel & 1\tabularnewline
\midrule
\textbf{Total} & \textbf{11}\tabularnewline
\bottomrule
\end{tabular}
\par\end{centering}
\caption{Composition of the STOC 2027 steering committee by country of affiliation
(\figref{sigact2})}
\end{table}

\begin{table}[H]
\begin{centering}
\begin{tabular}{lr}
\toprule 
\multicolumn{2}{c}{\textbf{SIGACT EC Ties to industry}}\tabularnewline
\midrule
No publicly-known recent AI/Big Tech ties: & 5\tabularnewline
Recent AI/Big Tech-backed funding: & 2\tabularnewline
Unmentioned current AI/Big Tech affiliation: & 1\tabularnewline
Mentioned current AI/Big Tech affiliation: & 3\tabularnewline
\midrule
\textbf{Total} & \textbf{11}\tabularnewline
\bottomrule
\end{tabular}
\par\end{centering}
\caption{Ties with industry within the STOC 2027 steering committee (\figref{sigact})}
\end{table}

\end{document}